\documentclass[12pt]{article}
\usepackage{amsmath}
\newcommand{\be}{\begin{equation}}
\newcommand{\ee}{\end{equation}}
\newcommand{\ba}{\begin{align}}
\newcommand{\ea}{\end{align}}

\begin{document}

\title{{\bf The Bekenstein Bound}}


\author{
Don N. Page
\thanks{Internet address:
profdonpage@gmail.com}
\\
Department of Physics\\
4-183 CCIS\\
University of Alberta\\
Edmonton, Alberta T6G 2E1\\
Canada
}

\date{2018 April 24}

\maketitle
\large
\begin{abstract}
\baselineskip 18 pt

	Bekenstein's conjectured entropy bound for a system of linear
size $R$ and energy $E$, namely $S \leq 2 \pi E R$, has counterexamples for
many of the ways in which the ``system,'' $R$, $E$, and $S$ may be
defined. One consistent set of definitions for these quantities
in flat Minkowski spacetime is that $S$ is the total von Neumann entropy
and $E$ is the expectation value of the energy in a ``vacuum-outside-$R$''
quantum state that has the the vacuum expectation values for all operators
entirely outside a sphere of radius $R$.  However, there are counterexamples
to the Bekenstein bound for this set of definitions.
Nevertheless, an alternative formulation ten
years ago by Horacio Casini for the definitions of $S$ and of $2 \pi E R$
have finally enabled a proof for this particular formulation of the Bekenstein bound. 

\end{abstract}
\normalsize
\baselineskip 16 pt
\newpage

\section{Introduction}

	Jacob Bekenstein was an outstanding scientist who made fundamental contributions to black hole thermodynamics and other related areas of theoretical physics.  He is perhaps best known for proposing that a black hole has an entropy proportional to its event horizon surface area $A$ and obeys a generalized second law, that the total black hole entropy plus the entropy of ordinary matter outside black holes never decreases \cite{Bekenstein:1972tm,Bekenstein:1973ur,Bekenstein:1974ax}.  This prediction was confirmed theoretically when Stephen Hawking discovered that black holes emit thermal radiation and have an entropy \cite{Hawking:1974rv,Hawking:1974sw,Hawking:1976de}
 \begin{equation}
 S_{\rm BH} = \frac{k c^3 A}{4 \hbar G} = \frac{A}{4},
 \label{eq:1}
 \end{equation}
where for the last quantity, and henceforth, I am using Planck units with $\hbar = c = G = k_\mathrm{Boltzmann} = 4\pi\epsilon_0 = 1$.  Note that the subscript ``BH'' on the entropy can mean either ``black hole'' or ``Bekenstein-Hawking,'' and this black hole entropy is indeed often called the Bekenstein-Hawking entropy.

	In this paper I shall focus on another interesting proposal that Jacob Bekenstein made that was motivated by the generalized second law but which exists independently for nongravitational systems, a conjectured upper bound of the entropy $S$ of a system of given energy $E$ and radius $R$ \cite{Bekenstein:1980jp}, known as the Bekenstein bound,
 \begin{equation}
 S \leq 2 \pi E R.
 \label{eq:1b}
 \end{equation}
Bekenstein and his colleagues developed many arguments and examples to support this conjecture 
\cite{Bekenstein:1980jp,Bekenstein:1981zz,Bekenstein:1982zv,Bekenstein:1982ph,Bekenstein:1983iq, Bekenstein:1984vm,Bekenstein:1987wf,Schiffer:1989et,Schiffer:1990hm,Bekenstein:1990du,Bekenstein:1993dz,Bekenstein:1999cf,Mayo:1999td,Bekenstein:1999bh,Bekenstein:2000ai,Bekenstein:2000sw, Bekenstein:2000ea,Bekenstein:2002cha,Bekenstein:2001qi,Bekenstein:2003dt,Bekenstein:2004sh, Bekenstein:2004ni}.
On the other hand, many counterarguments and counterexamples have also been put forward
\cite{Unwin:1982pg,Page:1982fj,Unruh:1982ic,Deutsch:1982zz,Unruh:1983ir,Ambjorn:1981xw,Radzikowski:1988kv,Unruh:1990hk,Pelath:1999xt,Wald:1999xu,Flanagan:1999jp,And,Wald:1999vt,Page:2000an,Page:2000up,Page:2000uq,Marolf:2003wu,Solodukhin:2000rt}.
Whether the conjectured bound (\ref{eq:1}) holds or not depends on what
systems are considered and how $R$, $E$, and $S$ are defined.

	Perhaps the simplest procedure \cite{Schiffer:1989et,Bekenstein:1990du} would be just to
consider quantum fields inside some bounded region within a sphere of
radius $R$ and put boundary conditions on the fields at the boundary of
the region.  However, this procedure leads to a large number of
counterexamples to Bekenstein's conjectured bound.  For example
\cite{Page:1982fj}, the Casimir effect can make $E<0$ for certain states of
quantum fields confined within a certain regions of radius $\leq R$,
violating the bound. If states with $E<0$ are excluded by definition,
one can still consider a mixed state with arbitrarily small positive $E$
that violates the bound. Even if $E$ is redefined to be the nonnegative
energy excess over that of the ground state \cite{Schiffer:1989et,Bekenstein:1990du}, one can
violate the bound by a mixed state that is almost entirely the ground
state and a tiny incoherent mixture of excited states, at least if the
entropy is defined to be $S = - tr\rho\ln\rho$ \cite{Deutsch:1982zz}. If $S$ is
instead defined to be $S = \ln n$ for a mixture of $n$ orthogonal pure
states (which would agree with $S = - tr\rho\ln\rho$ if the mixture had
equal probabilities $1/n$ for each of those $n$ pure states), then one
can violate the bound by an equal mixture of the ground state and the
first excited state of certain interacting fields with certain boundary
conditions that have the two lowest states nearly degenerate in energy
(separated by exponentially small tunnelling effects)
\cite{Page:1982fj,Page:2000an,Page:2000up}. If interacting fields are excluded from the
definition of allowable systems, one can get a violation by considering
a sufficiently large number $N$ of identical free fields, giving $n=N$
degenerate first excited states of finite energy but sufficiently large
entropy $S = \ln n = \ln N$ to violate (\ref{eq:1}) \cite{Page:1982fj,Unruh:1982ic}.
And even for a single free electromagnetic field, $S = \ln n$ can exceed
$2 \pi E R$ by an arbitrarily large factor by using boundary conditions
corresponding to an arbitrarily large number of parallel perfectly
conducting plates within the region of radius $R$ \cite{Page:2000an}, or by
using boundary conditions corresponding to an arbitrarily long coaxial
cable loop coiled up within the region \cite{Page:2000up}.

	However, other than in his papers with Schiffer \cite{Schiffer:1989et,Bekenstein:1990du},
Bekenstein generally advocated taking $E$ to be the total energy of
a complete system \cite{Bekenstein:1980jp,Bekenstein:1982zv,Bekenstein:1982ph,Bekenstein:1984vm,Bekenstein:2000sw,Bekenstein:2004sh}.  This would
disallow using just the energy of fields within a bounded region with
boundary conditions, since that would ignore the energy of whatever it
is that is providing the boundary conditions.  Therefore, all of the
counterexamples mentioned above would be excluded by this restriction. 
However, then the problem is to define what one means by the radius $R$
of the system.  In the flat spacetime case (quantum
fields in Minkowski spacetime) that we shall focus on here,
Bekenstein takes $R$ to mean the radius of a sphere which circumscribes
the system, which leaves the problem of what it means for a sphere to
circumscribe the complete system.

	In quantum field theory in Minkowski spacetime, the complete
system is the quantum state of the fields.  Since the quantum fields
extend all the way out to radial infinity, a sphere circumscribing the
entire system would have to be at $R=\infty$, which makes the Bekenstein
bound true (at least for states of positive energy and finite entropy)
but trivial.  To get a nontrivial bound, one needs to suppose that a
sphere of finite $R$ can circumscribe the system.  For example, one
might try to say that the sphere encloses all of the excitations of the
fields from the vacuum.  However, it is also hard to get this to occur
for a finite $R$.  For example, the wavefunction for any single particle
state that is a superposition of energy eigenstates of bounded energy
will not vanish outside any finite radius $R$, since a one-particle
wavefunction that does vanish outside a finite region must be a
superposition of arbitrarily large momentum components, which will have
unbounded energy.  Even if one looks at a composite system, such as a
hydrogen atom, and ignores the fact that its center of mass will have
amplitudes to be outside any finite sphere if it is made of purely
bounded energy components, the wavefunction for the relative position of
the electron and proton does not drop identically to zero outside any
finite separation distance for states that are superpositions of energy
eigenstates of bounded energy.  In particular, even if one fixed the
center of mass of a hydrogen atom in its ground state and ignored the
infinite energy from the resulting infinite uncertainty of the center of
mass momentum, the density matrix for the electron position would decay
only exponentially with distance from the center of mass and never go to
zero outside any sphere of finite radius $R$.

	Therefore, it is problematic to define the radius $R$ of a
sphere circumscribing a complete system in any quantum field theory. 
This issue was not addressed by Bekenstein and his collaborators,
but without such a definition, there is no nontrivial formulation of the
conjectured bound (\ref{eq:1}) for complete systems, but only its
trivial truth for any complete system with positive energy and finite
entropy that can only be circumscribed by the sphere enclosing all of
space, $R=\infty$.

	Here I shall first review ways I have proposed \cite{Page:2000uq} to define systems
and their radii $R$, energies $E$, and entropies $S$, so that for each,
there is a bound on $S$ for a given system as a function of finite $R$ and $E$.
These bounds do not have the form of Bekenstein's conjectured inequality
(\ref{eq:1}), though in some cases they may obey that inequality.

	Then I shall outline a new approach by Horacio Casini \cite{Casini:2008cr}
that gives new definitions for $S$ and for $2 \pi R E$, for which one can give
a rigorous proof of the Bekenstein bound.

\section{Vacuum-Outside-R States}

	The main new element in \cite{Page:2000uq} was the proposal to
define a system of radius $R$ (in flat Minkowski spacetime), not by
imposing boundary conditions on the field itself, but by imposing
conditions on the quantum state of the field so that outside a closed
ball of radius $R$ the quantum state is indistinguishable from the
vacuum at some time. Such a state will be called a vacuum-outside-$R$
state. (For simplicity, set this time to be $t=0$, and take the closed
ball, say $B$, to be the region $r\leq R$ on the $t=0$ hypersurface,
where $r$ is the standard radial polar coordinate giving the proper
distance from the coordinate origin on that hypersurface.) In other
words, a vacuum-outside-$R$ state of the system, say as expressed by its
density matrix $\rho$, is such that the expectation value of any
operator $O$ which is completely confined to the region $r>R$ when
written in terms of field and conjugate operators at $t=0$, is precisely
the same as the expectation value of the same operator in the vacuum
state $|0\!><\!0|$,
 \begin{equation}
 tr(O\rho) =\, <\!0|O|0\!>.
 \label{eq:2}
 \end{equation}
In particular, all the $n$-point functions for the field and for its
conjugate momentum in the state $\rho$ are the same as in the vacuum
state, if all of the $n$ points are outside the ball of radius $R$ and
on the hypersurface $t=0$. Of course, the $n$-point functions need not
be the same as their vacuum values if some or all of the points are
inside the ball.

	If operators confined to the three-dimensional region $r>R$ and
$t=0$ (say $C$, to give a name to this achronal spacelike surface, the
$t=0$ hypersurface with the central closed ball $B$, $r\leq R$,
excluded) have the same expectation value in the vacuum-outside-$R$
state as in the vacuum state, the same will be true in any quantum field
theory that I shall call ``strongly causal'' for all operators confined
to the Cauchy development or domain of dependence \cite{Hawking:1973uf} of $C$, the
larger four-dimensional region $r>R+|t|$ (say $D$) that is the set of
all points in the Minkowski spacetime such that every inextendible
(endless) causal, or non-spacelike (everywhere timelike or lightlike),
curve through such a point intersects the partial Cauchy surface $C$.
Just as solutions of hyperbolic wave equations in $D$ are determined by
the data on $C$, so the part of the quantum state of a strongly causal
field in $D$, as represented by the expectation values of operators
confined to $D$, is determined by the part of the quantum state in $C$,
as represented by the expectation values of operators confined to $C$.
(For some interacting quantum field theories, the expectation values of
operators confined to the three-dimensional spacelike surface $C$ may be
too ill-defined for these theories to be ``strongly causal'' in my
sense, but a wider class of these theories may be ``weakly causal'' in
the sense that sufficiently many operators smeared over, but confined
to, an arbitrarily thin-in-time four-dimensional slab, say $E$,
containing $C$ within $D$, have well-defined expectation values that
determine the expectation values of all operators smeared over, but
confined to, any part of $D$.)

	Henceforth I shall restrict attention to strongly causal and
weakly causal quantum field theories, calling them simply causal quantum
field theories for short. I shall also assume that any quantum field theory under
consideration is a nongravitational Lorentz-invariant quantum field
theory in Minkowski spacetime, and that it has a unique pure state of
lowest Minkowski energy $E = 0$ (the expectation value of the
Hamiltonian $H$ that generates translations in the time coordinate $t$
in some Lorentz frame, with the arbitrary constant in the Hamiltonian
being adjusted to give the lowest energy state zero energy).

	Therefore, for such a causal nongravitational quantum field
theory in Minkowski spacetime, I proposed \cite{Page:2000uq} that the radius $R$ be
defined so that all of the operators constructed from field and
conjugate momentum operators smeared over regions confined to the region
$D$, $r>R+|t|$ in some Lorentz frame, have in the particular quantum
state being considered (a vacuum-outside-$R$ state) the same expectation
values that they have in the vacuum state for that quantum field theory.
The energy $E$ of the state $\rho$ can then be simply defined to be the
expectation value,
 \begin{equation}
 E \equiv tr(H\rho),
 \label{eq:3}
 \end{equation}
of the Hamiltonian $H$ that generates time translations in the same
Lorentz frame. Because the energy $E$ has been defined to have the
minimum value of zero for the unique pure vacuum state, there is no
problem here with negative Casimir energies. In other words, the energy
is that of the complete system over all of Minkowski spacetime.

	Obviously we would also like a definition of
the entropy $S$ that has a minimum value of zero,
which it should attain for the pure vacuum state.
One simple definition is the von Neumann entropy,
 \begin{equation}
 S = S_{\rm vN} \equiv - tr\rho\ln\rho,
 \label{eq:4}
 \end{equation}
using the density matrix $\rho$ for the full state
of the quantum field, over the entire Minkowski spacetime.

\section{Entropy Bounds for Vacuum-Outside-R States}

	Now we may conjecture that for any vacuum-outside-$R$ state of
any particular causal nongravitational quantum field theory
in Minkowski spacetime, one which has the vacuum expectation
values in the region $D$, $r>R+|t|$ (the region causally
disconnected from the ball $r \leq R$ at $t = 0$),
the von Neumann entropy is bounded above by some function
$\sigma_{\rm vN}$ (depending on the quantum field theory in question)
of the radius $R$ and energy expectation value $E$:
 \begin{equation}
 S_{\rm vN} \leq \sigma_{\rm vN}(R,E).
 \label{eq:5}
 \end{equation}
Define this function $\sigma_{\rm vN}(R,E)$ to be the
least upper bound on the von Neumann entropy
of any state which is vacuum outside the radius $R$
and which has energy expectation value $E$.
 
	In the case of a scale-invariant quantum field,
such as a free massless field, or say a massless scalar
field $\phi$ with a $\lambda \phi^4$ self-coupling potential,
the least upper bound function $\sigma_{\rm vN}(R,E)$
will actually be a function of the single dimensionless
variable
 \begin{equation}
 x \equiv 2 \pi R E,
 \label{eq:6}
 \end{equation}
say
 \begin{equation}
 \sigma_{\rm vN}(R,E) = \sigma_{\rm N}(x).
 \label{eq:7}
 \end{equation}

	Bekenstein's conjectured entropy bound (\ref{eq:1}),
if $R$, $E$, and $S$ were defined as done above,
would be $\sigma_{\rm vN}(R,E) \leq x$,
whether or not the quantum field theory
is scale invariant, or
 \begin{equation}
 B_{\rm vN}(R,E) \equiv \frac{\sigma_{\rm vN}(R,E)}{x}
                 \equiv \frac{\sigma_{\rm vN}(R,E)}{2\pi R E} \leq 1.
 \label{eq:7b}
 \end{equation}
If the quantum field theory is scale invariant,
we can define
 \begin{equation}
 B_{\rm N}(x) \equiv \frac{\sigma_{\rm N}(x)}{x},
 \label{eq:7c}
 \end{equation}
which should also be less than or equal to unity
if Bekenstein's bound applies.

	For a set of one or more free massless fields
and vacuum-outside-$R$ states with $x \gg 1$,
one would expect that the highest entropy would
be given by a mixed state that at $t=0$ is approximately
a high-temperature ($RT \gg 1$) thermal radiation state
for $r < R$, surrounded by vacuum for $r > R$.
A high-temperature thermal radiation state has an
energy density for massless fields of approximately $a_r T^4$,
and hence an entropy density $(4/3)a_r T^3$, where
 \begin{equation}
 a_r = \frac{\pi^2}{30}(n_b + \frac{7}{8}n_f)
 \label{eq:8}
 \end{equation}
is the radiation constant for $n_b$ independent
bosonic degrees of freedom for each momentum
(e.g., $n_b$ different spin or helicity states)
and for $n_f$ fermionic degrees of freedom.
Therefore, in this case with $x \gg 1$,
 \begin{equation}
 B_{\rm vN}(R,E) = \frac{\sigma_{\rm N}(x)}{x}
 \approx \left(\frac{2^7 a_r}{3^5 \pi^2 x} \right)^\frac{1}{4}
 = \left[\frac{2^6}{3^6 5 x}(n_b + \frac{7}{8}n_f)\right]^\frac{1}{4},
 \label{eq:9}
 \end{equation}
which is indeed less than 1, thus obeying Bekenstein's
conjectured bound, for
 \begin{equation}
 x \geq \frac{2^7 a_r}{3^5 \pi^2}
   = \frac{2^6}{3^6 5}(n_b + \frac{7}{8}n_f)
   = \frac{64 n_b + 56 n_f}{3645},
 \label{eq:10}
 \end{equation}
if $x$ is also large enough that Eq.\ (\ref{eq:9}) is a good
approximation. Thus one would expect that Bekenstein's conjectured
bound, using the definitions above for $R$, $E$, and $S$, holds for a
fixed set of free massless quantum fields at sufficiently large $x
\equiv 2 \pi R E$.

	On the other hand, the definitions above
for $R$, $E$, and $S$ still permit Bekenstein's conjectured
bound applied to them to be violated for sufficiently small $x$,
as we can see by the following construction:

	A way to construct vacuum-outside-$R$ states,
quantum states of a free quantum field
theory that have vacuum expectation values in the region $D$,
$r>R+|t|$, is to apply to the vacuum state unitary operators
constructed from fields and/or conjugate momenta smeared within
the region $r<R$ at $t=0$.
In particular, if $h$ is an hermitian operator
constructed from fields and/or conjugate momenta
smeared within $r<R$ at $t=0$, then $U = e^{ih}$
is such a unitary operator, and
 \begin{equation}
 |\psi\!> = U|0\!> = e^{ih}|0\!>
 \label{eq:11}
 \end{equation}
is a pure quantum state that has precisely the vacuum expectation
values in the region $D$.  This result can be seen formally from
the fact that any operator $O$ confined to the region $D$
(the four-dimensional region $r>R+|t|$)
that is causally disconnected from the ball $B$
(the three-dimensional region $r\leq R$ on the $t=0$ hypersurface)
commutes with the operators $h$ and $U$ that are confined
to that hypersurface, $[O,h] = [O,U] = 0$, so
 \begin{equation}
 tr(O\rho) = <\!\psi|O|\psi\!>
           = <\!0|U^{-1}OU|0\!> = <\!0|U^{-1}UO|0\!> = <\!0|O|0\!>.
 \label{eq:12}
 \end{equation}
 
	If $\{h_i\}$ is a set of hermitian operators
that each are confined to the ball $B$ (i.e., are constructed
from fields and momenta that are smeared only over that region),
and if $\{q_i\}$ is a set of positive numbers that sum to unity, then
 \begin{equation}
 \rho = \sum_i q_i e^{i h_i}|0\!><\!0|e^{-i h_i}
 \label{eq:13}
 \end{equation}
is a more general vacuum-outside-$R$ state,
since this density matrix gives vacuum expectation values,
$tr(O\rho) = <\!0|O|0\!>$, for any operator $O$ confined to the region
$D$ that is causally disconnected from $B$
(i.e., having no causal curves, either timelike or lightlike,
intersecting both the ball $B$ of $r\leq R$ at $t=0$
and the region $D$ with $r>R+|t|$).

	Here let us consider the simple example in which
$i$ takes only the two values 1 and 2, and $h_1=0$ and $h_2=h$.
Let $q_1=1-q$ and $q_2=q$, and let
 \begin{equation}
 e^{ih}|0\!> = U|0\!> = |\psi\!> = c|0\!>+s|1\!>
 \label{eq:14}
 \end{equation}
in terms of a decomposition of $|\psi\!>$ into the two
orthonormal states $|0\!>$ and
$|1\!>=(|\psi\!>-<\!0|\psi\!>|0\!>)/\sqrt{1-|\!\!<\!0|\psi\!>\!\!|^2}$, 
so
 \begin{equation}
 c = <\!0|\psi\!> = <\!0|U|0\!> = <\!0|e^{ih}|0\!>,
 \label{eq:15}
 \end{equation}
 \begin{equation}
 s = \sqrt{1-|\!\!<\!0|U|0\!>\!\!|^2}
   = \sqrt{1-|c|^2}.
 \label{eq:16}
 \end{equation}
Note that $|1\!><\!1|$ by itself is not generically
a vacuum-outside-$R$ state.

	Now Eq.\ (\ref{eq:13}) gives the density matrix as
 \begin{eqnarray}
 \rho &=& (1-q)|0\!><\!0|+q|\psi\!><\!\psi| \nonumber \\
      &=& (1-qs^2)|0\!><\!0|+qcs|0\!><\!1|
      +q\bar{c}s|1\!><\!0|+qs^2|1\!><\!1|,
 \label{eq:17}
 \end{eqnarray}
a density matrix in the two-dimensional space
of pure states spanned by the two orthonormal pure
states $|0\!>$ and $|1\!>$.  The two eigenvalues of
this density matrix are, say, $p$ and $1-p$
(since their sum is $tr\rho=1$), with product
 \begin{eqnarray}
 y \equiv p(1-p) &=& \frac{1}{2}\{[p+(1-p)]^2-[p^2+(1-p)^2]\}
        = \frac{1}{2}\{[tr(\rho)]^2-[tr(\rho^2)]\} \nonumber \\
	&=& q(1-q)(1-|\!\!<\!0|U|0\!>\!\!|^2) = q(1-q)s^2.
 \label{eq:18}
 \end{eqnarray}

	The expectation value of the energy of this
mixed state is, since I have assumed $H|0\!>=0$,
 \begin{equation}
 E = tr(H\rho) = q<\!\psi|H|\psi\!> = q<\!0|U^{-1}HU|0\!>.
 \label{eq:19}
 \end{equation}
Then
 \begin{equation}
 x \equiv 2 \pi R E = 2 \pi R q<\!0|U^{-1}HU|0\!>.
 \label{eq:20}
 \end{equation}

	The von Neumann entropy of this mixed state is
 \begin{eqnarray}
 S_{\rm vN}(y) &=& - tr(\rho\ln\rho) = -p\ln{p}-(1-p)\ln{(1-p)}
  		 \nonumber \\
               &=& \frac{y}{\frac{1}{2}(1+\sqrt{1-4y})} \ln{\frac{1}{y}}
	        + \sqrt{1-4y} \ln\frac{1}{\frac{1}{2}(1+\sqrt{1-4y})}
		 \nonumber \\
	       &\approx& y[(1+y+2y^2)\ln\frac{1}{y}
	        +(1-\frac{1}{2}y-\frac{5}{3}y^2)],	
 \label{eq:21}
 \end{eqnarray}
a monotonically increasing function of $y\equiv q(1-q)s^2 \leq 1/4$,
where the last approximate equality of Eq.\ (\ref{eq:21}) applies for 
$y \ll 1$.

	As $q$ and/or $h$ is reduced toward zero,
$x$, $y$, and $S$ also decrease toward zero, but whereas
$x$ and $y$ asymptotically decrease linearly with $q$,
the dominant term of
$S$ has an extra logarithmic factor that grows with
the reduction of $y$, so the ratio,
 \begin{equation}
 B \equiv \frac{S_{\rm vN}}{x} \equiv \frac{S_{\rm vN}}{2\pi R E}
    \approx \frac{y}{x}\left(\ln\frac{1}{y} + 1 \right)
 \label{eq:22}
 \end{equation}
when $y \ll 1$,
increases without limit as $y$ is reduced toward zero.
Therefore, when $y$ is made sufficiently small
(e.g., by making $q$ sufficiently small),
Bekenstein's conjectured bound is violated for the definitions of
$R$, $E$, and $S$ used here.

	I then went on \cite{Page:2000uq} to give some examples in the free quantum field theory of a single massless scalar field, finding for a particular set of vacuum-outside-$R$ states that
 \begin{equation}
 B \sim \frac{1}{\gamma}\left(\ln\frac{\gamma}{x} + 1 \right),
 \label{eq:44b}
 \end{equation}
with numerical calculations giving
 \begin{equation}
 \gamma \approx 12.6048,
 \label{eq:55b}
 \end{equation}
thus violating the Bekenstein bound, $B \leq 1$, for
 \begin{equation}
 x \equiv 2 \pi R E < \gamma e^{1-\gamma} \approx 0.000115.
 \label{eq:56b}
 \end{equation}

	I further conjectured \cite{Page:2000uq} that the true upper bound has the same asymptotic form with the same value of $\gamma$,
 \begin{equation}
 B_{\rm N}(x) \equiv \frac{\sigma_{\rm N}(x)}{x} \sim \frac{1}{\gamma}\left(\ln\frac{\gamma}{x} + 1 \right).
 \label{eq:58b}
 \end{equation}

\section{A Provable Formulation of the Bekenstein Bound}

	Another approach developed by Raphael Bousso \cite{Bousso:2003cm} is to try to bound the size of a system by including interactions.  This goes some ways toward giving a proof of the Bekenstein bound, though I am a bit confused as to how interactions could possibly bind nonzero energy eigenstates into structures that are strictly zero outside some finite distance (even if this distance is just a relative distance from the center of mass in a zero-momentum eigenstate, as was done to avoid the problem of confining the center of mass to a finite region, which would, as discussed above, lead to components of the quantum state having arbitrarily large momenta and hence components with arbitrarily large energy).  However, this paper does not seem to give precise upper bounds on the entropy as a function of the energy and size of the system.
	
	An approach that finally does seem to allow one to get a proof of at least one form of the Bekenstein bound is one developed by Horacio Casini \cite{Casini:2008cr}, building upon previous work \cite{Marolf:2003sq,Marolf:2004et,Wald:1999xu}.
	
	This paper \cite{Casini:2008cr} showed some awareness of my paper \cite{Page:2000uq} by citing it and mentioning its definition of the localization of a system (vacuum-outside-$R$ states), but  \cite{Casini:2008cr} incorrectly assumed \cite{Page:2000uq} defined the entropy by using the microcanonical ensemble for the number of such states below energy $E_0$, which does not work because such states do not form a vector space.  However, the erroneous assumption that I did not have a good way to calculate the entropy may have been fortunate, because it does seem rather hard to make progress toward finding a way to prove, rather than disprove, the Bekenstein bound using vacuum-outside-$R$ states, so the erroneous assumption perhaps prevented Casini from getting bogged down in this difficult approach and may have helped motivate him to seek a more fruitful direction for finding at least one way in which the Bekenstein bound seems to be true.
	
	Casini \cite{Casini:2008cr} takes quantum field theory in the right Rindler wedge of flat Minkowski spacetime, $z > |t|$, with Cauchy surface the spatial half-plane $V$ defined by $z > 0$ at $t=0$.  A quantum field theory state over the entire spatial plane $t=0$ then leads to a mixed state $\rho_V$ on the half-plane and its Cauchy development.  Because of the sharp boundary of the half-plane at $z=0$, the von Neumann entropy of $\rho_V$, namely
 \begin{equation}
 S(\rho_V) = -{\rm tr}\rho_V \ln{\rho_V},
 \label{eq:59b}
 \end{equation}
diverges.  However, if one subtracts the von Neumann entropy of the state $\rho_V^0$ that is the restriction of the global vacuum state to the spatial half-plane, one gets the finite result
 \begin{equation}
 S_V = S(\rho_V) - S(\rho_V^0).
 \label{eq:60b}
 \end{equation}
 
	The ingenious proposal of Casini \cite{Casini:2008cr} is to replace $2\pi E R$ by 
 \begin{equation}
 K_V = {\rm tr}(K\rho_V) - {\rm tr}(K\rho_V^0),
 \label{eq:61b}
 \end{equation}
where
 \begin{equation}
 K = 2\pi \int dx\, dy \int_0^\infty z\, \mathcal{H}(x,y,z,0),
 \label{eq:62b}
 \end{equation}
is the boost operator, with $\mathcal{H}(x,y,z,t)$ being the Hamiltonian density operator.

	Then with these clever identifications of what the quantities $S$ and $2\pi E R$ should be, the Bekenstein bound becomes $S_V \leq K_V$, or
 \begin{equation}
 S(\rho_V) - S(\rho_V^0) \leq {\rm tr}(K\rho_V) - {\rm tr}(K\rho_V^0).
 \label{eq:63b}
 \end{equation}
However, the state that is the restriction of the global vacuum state to the spatial half-plane is
 \begin{equation}
 \rho_V^0 = \frac{e^{-K}}{{\rm tr}\,e^{-K}},
 \label{eq:64b}
 \end{equation}
so $K = -\ln{\rho_V^0} -\ln{{\rm tr}\,e^{-K}}$, and one gets (using ${\rm tr}\,\rho_V = {\rm tr}\,\rho_V^0 = 1$) that the Bekenstein bound is equivalent to the known nonnegativity of the relative entropy \cite{Wehrl:1978zz,Vedral:2002zz,Nielsen-Chuang},
 \begin{equation}
 S(\rho_V|\rho_V^0) \equiv {\rm tr}(\rho_V\ln{\rho_V}) - {\rm tr}(\rho_V\ln{\rho_V^0}) \ge 0.
 \label{eq:65b}
 \end{equation}
 
 	These arguments \cite{Casini:2008cr} were made in terms of a theory with a cutoff, but Robert Longo and Feng Xu \cite{Longo:2018zib} were able to recast them in full quantum field theory, by operator algebraic methods, into a rigorous proof of the Bekenstein bound \cite{Bekenstein:1980jp}. 
	
	One disadvantage of Casini's form of the Bekenstein bound as the inequality (\ref{eq:63b}), or $S_V \leq K_V$, is that both sides of the inequality can be negative \cite{Blanco:2013lea}, unlike the entropy $S$ and energy $E$ in the vacuum-outside-$R$ formulation discussed earlier above.  However, Blanco and Casini \cite{Blanco:2013lea} have also given an alternative form of the Bekenstein bound in which both the expressions for the entropy $S$ and for $2\pi E R$ are nonnegative.

\section{Conclusions and Acknowledgments}

	In conclusion, we have found that one can formulate precise
definitions for entropy bounds of a complete quantum field system
over all of Minkowski spacetime by giving precise definitions 
for the size $R$ of the system on the hypersurface $t = 0$. 
In particular, $R$ may be defined for vacuum-outside-$R$ states
as the largest round two-sphere on the $t=0$ hypersurface, 
outside of which all of the operators have the same expectation values
as in the vacuum state.  However, this definition, plus the definition
of $E$ as the expectation value of the Hamiltonian and 
$S$ as the von Neumann entropy of the entire quantum state,
leads to violations of the Bekenstein bound.
Nevertheless, an alternative set of definitions for $S$
and for $2 \pi R E$ by Horacio Casini have finally succeeded
in leading to a rigorous proof of one form of the Bekenstein bound.

	This work was supported by the
Natural Sciences and Engineering Research Council of Canada.


\end{document}